# Theory of thermally activated vortex bundles flow over the directional-dependent potential barriers in type-II superconductors

Wei Yeu Chen

*Department of Physics, Tamkang University, Tamsui 25137, Taiwan, R.O.C.*

## Abstract

The thermally activated vortex bundle flow over the directional-dependent energy barrier in type-II superconductors is investigated. The coherent oscillation frequency and the mean direction of the random collective pinning force of the vortex bundles are evaluated by applying the random walk theorem. The thermally activated vortex bundle flow velocity is obtained. The temperature- and field-dependent Hall and longitudinal resistivities induced by the bundle flow for type-II superconducting bulk materials and thin films are calculated. All the results are in agreement with the experiments.

Tel.: 886-2-28728682; fax: 886-2-28728682

E-mail: wychen@mail.tku.edu.tw (W.Y. Chen)

## 1. INTRODUCTION

In type-II superconductor [1-24], when the apply magnetic field $B > B_{C_1}$, the field penetrate the superconductor to form a long-range order of vortex lattice or flux line lattice when the material is homogeneous [1]. However, quenched disorder always destroys the long-range order of the vortex lattice, after which only short-range order or the vortex bundle remains [3, 6-11]. The vortex lines inside the vortex bundle oscillate about their equilibrium positions due to thermally agitation for finite temperature [7, 8].

In this paper we are going to develop a self-consistent theorem of thermally activated vortex bundles flow over the directional-dependent potential barriers. The basic physical idea is that the directional-dependent potential barrier of the vortex bundle for calculating the bundle flow velocity, actually contains the bundle flow velocity itself. The directional-dependent energy barrier means that the energy barrier is a function of direction of the thermally activated motion. The coherent frequency of oscillation of the vortex bundle, and the mean direction of the random collective pinning forces of the vortex bundles are evaluated by applying the theorem of random walk. The bundle flow velocity is then obtained self-consistently. Finally, the Hall and longitudinal resistivities generated by the bundle flow are calculated.

The rest of this paper is organized as follows. In section 2, a mathematical model is presented. In section 3, the coherent frequency of oscillation the vortex bundles and the mean direction of the random collective pinning forces of the vortex bundles are calculated. The directional-dependent energy barrier generated by the Magus force, the random collective pinning force and strong pinning force of the vortex bundle is obtained in section 4. In section 5, the bundle flow velocity is evaluated self-consistently, the Hall and longitudinal resistivities generated by the bundle flow are calculated. Finally, the concluding remarks are given in section 6.

## 2. MATHEMATICAL DESCRIPTION OF THE MODEL

Let us consider a type-II conventional or high-$T_c$ superconductor, the Hamiltonian of the fluctuation for the flux line lattice (FLL) in the $z-direction$ is given by [9, 11-12]

$$H = H_f + H_R \tag{1}$$

where $H_f = H_{kin} + H_e$ presents the Hamiltonian for the free modes [9, 11-12], with $H_{kin}$ is the kinetic energy part [9, 11-12]

$$H_{kin} = \frac{1}{2\rho} \sum_{\vec{K}\mu} P_\mu(\vec{K}) P_\mu(-\vec{K}) \tag{2}$$

$H_e$ the elastic energy part [9, 11-13],

$$H_e = \frac{1}{2} \sum_{\vec{K}\mu\nu} C_L K_\mu K_\nu S_\mu(\vec{K}) S_\nu(-\vec{K}) + \frac{1}{2} \sum_{\vec{K}\mu} (C_{66} K_\perp^2 + C_{44} K_z^2) S_\mu(\vec{K}) S_\mu(-\vec{K}) \tag{3}$$

and $H_R$ represents the random Hamiltonian, given as [9, 11-12],

$$H_R = \sum_{\vec{K}\mu} f_{R\mu}(\vec{K}) S_\mu(-\vec{K}) \tag{4}$$

where $(\mu, \nu) = (x, y)$, $\rho$ is the effective mass density of the flux line [14], $K_\perp^2 = K_x^2 + K_y^2$, $P_\mu(\vec{K}), S_\mu(\vec{K})$ are the Fourier transformations of the momentum and displacement operators, $C_L, C_{11}, C_{44}$ and $C_{66}$ are temperature- and $\vec{K}$ -dependent bulk modulus, compression modulus, tilt modulus and shear modulus, respectively. While $\vec{f}_R(\vec{K})$ is the Fourier transformation of the collective pinning force $\vec{f}_R(\vec{r}) = -\vec{\nabla} V_R(\vec{r})$, with $V_R(\vec{r})$ the random potential energy of the collective pinning [15-16], which is the sum of the contributions of defects within a distance $\xi$ away from the vortex core position $\vec{r}$, where $\xi$ is the temperature-dependent coherent length. The correlation functions of the random collective pinning force are assumed to be the short-range correlation [9],

$$\overline{\langle\langle f_{R\alpha}(\vec{k}) f_{R\beta}^*(\vec{k}') \rangle\rangle_{th}} = \beta^C(T, B)\, \delta_{\alpha\beta}\, \delta(\vec{k} - \vec{k}') \tag{5}$$

where $\overline{\langle\langle \quad \rangle\rangle_{th}}$ are the quantum, thermal, and random averages, and $\beta^C(T, B)$ is the temperature- and magnetic field-dependent correlation strength.

The displacement operator $S_\mu(\vec{K})$ is given by

$$S_\mu(\vec{K}) = S_{R\mu}(\vec{K}) + S_{f\mu}(\vec{K}) \tag{6}$$

with $S_{R\mu}(\vec{K})$ denotes the deformation displacement operator of the FLL due to the collective pinning of the random function $\vec{f}_{R\mu}(\vec{K})$, and $S_{f\mu}(\vec{K})$ is the displacement operator for the fluctuation of the free modes. They are given by

$$S_{R\mu}(\vec{K}) = [(\vec{K} \cdot \vec{f}_R(\vec{K})) \frac{\delta_{\alpha,1}}{K_\perp}] \cdot \frac{1}{C_{11} K_\perp^2 + C_{44} K_z^2} + [f_{R\alpha}(\vec{K}) - (\vec{K} \cdot \vec{f}_R(\vec{K})) \frac{\delta_{\alpha,1}}{K_\perp}] \cdot \frac{1}{C_{66} K_\perp^2 + C_{44} K_z^2} \tag{7}$$

$$S_{f\mu}(\vec{K}) = \sqrt{\frac{\hbar}{2\rho\omega_{K\mu}}}(\alpha^{+}_{-\vec{K}\mu} + \alpha_{\vec{K}\mu}) \tag{8}$$

respectively, where $\mu = 1$ presents the component parallel to the $\vec{K}_{\perp}$ direction, while $\mu = 2$ is perpendicular to the $\vec{K}_{\perp}$ direction. It is understood that the free Hamiltonian can be diagonalized with the eigenmodes spectrum [9, 11-12]

$$\omega_{K1} = [\frac{1}{\rho}(C_{11}K_{\perp}^2 + C_{44}K_z^2)]^{\frac{1}{2}}$$

$$\omega_{K2} = [\frac{1}{\rho}(C_{66}K_{\perp}^2 + C_{44}K_z^2)]^{\frac{1}{2}} \tag{9}$$

with $\alpha^{+}_{\vec{K}\mu}$, $\alpha_{\vec{K}\mu}$ are the creation and the annihilation operators for the corresponding eigenmodes.

The quenched disorder destroys the long-range order of the FLL, after which only short-range order, the vortex bundle, prevails. The corresponding size of vortex bundle $|\vec{R}|$ is determined by the relation [11-12]

$$\overline{<<|\vec{S}_R(\vec{R}) - \vec{S}_R(0)|^2>>_{th}} = r_f^2 \tag{10}$$

where $r_f$ represents the random collective pinning force range.

## 3. COHERENT OSCILLATION FREQUENCY AND MEAN DIRECTION OF COLLECTIVE PINNING FORCE FOR VORTEX BUNDLE

Let us consider a p-type superconductor, with the applied magnetic field $\vec{B}$ in the z-axis and the external current density in the x-axis $\vec{J} = J\,\vec{e}_x$ with $J < J_C$ where $J_C$ is the critical current density of the superconductor. The equation of motion of the vortex line inside the vortex bundle driving by the thermal radiation of frequency $\omega$ is

$$\frac{d^2\vec{r}_l}{dt^2} = \frac{q\vec{E}e^{i\omega t}}{M_v} + \frac{J\Phi_0}{M_v}\vec{e}_x \times \vec{e}_z - (\frac{1}{\tau_R}\frac{d\vec{r}_v}{dt}) - (\frac{k_R}{M_v}\vec{r}_v) + \frac{\vec{f}_{el}}{M_v} \tag{11}$$

where $\vec{r}_v$ is the displacement of vortex line, $\vec{E}$ denotes the electric field of the thermal radiation, $M_v$ stands for the effective mass, and $q$ is the total circulating charge of the vortex line, $\Phi_0$ is the unit flux, $1/\tau_R$ characterizes the damping rate associated with the motion of the vortex line, $k_R$ represents the restoring force constant for the vortex line under the action of random collective pinning force, and $\vec{f}_{el}$ is the elastic force of the vortex line inside the vortex bundle. The homogeneous solution of the equation (11) vanishes quickly due to the presence of damping. The particular solution includes two parts: the time-dependent and time-independent parts.

The time-dependent part of the particular solution oscillates with frequency $\omega$ about a equilibrium position, which is determined by the time-independent part of the particular solution. By identifying the oscillation energy of the vortex line inside the potential barrier with the thermal energy, the thermal oscillation frequency $\nu$ of the individual vortex inside the potential barrier can be expressed as

$$\nu = \bar{\nu}\sqrt{T} \tag{12}$$

with the proportional constant $\bar{\nu} = (1/\pi A)\sqrt{k_B / 2M_v}$ , $A$ stands for the random and thermal averages amplitude of the oscillation of the vortex line in the bundle, $k_B$ is the Boltzmann constant.

However, the oscillations of vortex lines inside the vortex bundle are not coherent, namely, their oscillations are at random. To obtain the coherent oscillation frequency $\nu_c$ of the vortex bundle as a whole, by utilizing the random walk's theorem, the frequency $\nu$ in equation (12) must be divided by the square root of N, the number of vortices inside the vortex bundle

$$\nu_c = \frac{\nu}{\sqrt{N}} = \frac{\bar{\nu}\sqrt{T}\sqrt{\Phi_0}}{R\sqrt{\pi B}} \tag{13}$$

where $R$ is the transverse size of the vortex bundle and $B$ is the value of the applied magnetic field.

The time-independent part of the particular solution is give by

$$\vec{r}_p = \frac{-J\,\Phi_0\,\vec{e}_y + \vec{f}_{el}}{k_R} \tag{14}$$

This result indicates that the vortex line moves to a new equilibrium position $\vec{r}_p$ from its original one. Since the elastic force is much less than the Lorentz force, the angle between the random collective pinning force and the positive y-direction measured in the counterclockwise sense can be obtained approximately as

$$\theta \cong \frac{|\vec{f}_{el}|}{|\vec{f}_L|} = \frac{|\vec{f}_{el}|}{J\Phi_0} \tag{15}$$

where $|\vec{f}_{el}|$ and $|\vec{f}_L|$ are the magnitudes of elastic force and Lorentz force of the vortex line. Taking into account the fact that the compression modulus $C_{11}$ is much larger than shear modulus $C_{66}$ [9-10], owing to the thermal fluctuations, the magnitude of the displacement vector $|\vec{S}_f(\vec{r})|$ of vortex line inside the vortex bundle as well as its corresponding magnitude of the elastic force $|\vec{f}_{el}|$ is proportional to $\sqrt{k_B/C_{66}}$ , hence,

$$|S_f| \sim |\vec{f}_{el}| \sim \sqrt{k_B/C_{66}} \sim (1/\sqrt{B})\sqrt{\frac{T}{T_C - T}} \quad (16)$$

The temperature- and field-dependent $\theta$ can therefore be written as

$$\theta(T,B) = \alpha \frac{1}{\sqrt{B}}\sqrt{\frac{T}{T_c - T}} \quad (17)$$

$\alpha$ is a proportional constant. By applying the random walk theorem, the mean angle $\Theta\,(T,B)$ between the random collective pinning force of vortex bundle and positive y-direction measures in counterclockwise sense, can be expressed as

$$\Theta(T,B) = \sqrt{N}\,\theta(T,B) = \overline{\alpha}\,\sqrt{\frac{T}{T_c - T}} \quad (18)$$

where $\overline{\alpha} = \alpha\, R\sqrt{\pi/\Phi_0}$ .

## 4. DIRECTIONAL-DEPENDENT ENERGY BARRIER

In this section we shall calculate the directional-dependent energy barrier of the vortex bundles formed by the Magnus force, random collective pinning force, together with the strong pinning force inside the vortex bundle for the applied magnetic field in the z-direction $\vec{B} = B\vec{e}_z$ and transport current density in the x-direction $\vec{J} = J\,\vec{e}_x$, with $J < J_C$ . Let the velocity of the thermally activated vortex bundle flow be $\vec{v}_b$, and the Magnus force acting on the vortex bundle is given by

$$\vec{F}_M = \overline{V}\; n_s\; e\; (\vec{v}_T - \vec{v}_b) \times B\,\hat{e}_z \quad (19)$$

where $\overline{V}$ is the volume of the vortex bundle, $e$ is the electron charge, $n_s$ and $\vec{v}_T$ are the supercharge density and its velocity, respectively, with $\vec{J} = n_s e \vec{v}_T = J\,\vec{e}_x$ . From the theory of mechanics, the potential generated by a force field $\vec{F}(r)$ is

$$V(\vec{R}) - V(0) = -\int_0^{\vec{R}} \vec{F}(r)\cdot d\,\vec{r} \quad (20)$$

Let us assume that the potential barrier generated by the strong pinning force due to the randomly distributed strong pinning sites inside the vortex bundle be $U$ . After some algebra, the directional-dependent energy barrier of the vortex bundles both in the positive and negative x-direction as well as the positive and negative y-direction are obtained as

$$U+\overline{V}R(JB\frac{v_{by}}{v_T}-<F_{p_x}>_R)$$

$$U-\overline{V}R(JB\frac{v_{by}}{v_T}-<F_{p_x}>_R)$$

$$U+\overline{V}R(JB-JB\frac{v_{bx}}{v_T}-<F_{p_y}>_R)$$

$$U-\overline{V}R(JB-JB\frac{v_{bx}}{v_T}-<F_{p_y}>_R) \tag{21}$$

respectively, once again, $R$ represents the transverse size of the vortex bundle, the range of $U$ is assumed to be the order of $R$ , and $<\vec{F}_p>_R$ stands for the random average of the random collective pinning force per unit volume.

## 5. BUNDLE FLOW VELOCITY AND ITS INDUCED LONGITUDINAL AND HALL RESISTIVITIES

The self-consistent equations for the velocity of thermally activated vortex bundles flow over the directional-dependent energy barrier are therefore obtained in components as

$$v_{by}=\nu_c R\{\exp[\frac{-1}{k_BT}(U+\overline{V}R(JB-JB\frac{v_{bx}}{v_T}-<F_{p_y}>_R))] \\ -\exp[\frac{-1}{k_BT}(U-\overline{V}R(JB-JB\frac{v_{bx}}{v_T}-<F_{p_y}>_R))]\} \tag{22}$$

$$v_{bx}=\nu_c R\{\exp[\frac{-1}{k_BT}(U+\overline{V}R(JB\frac{v_{by}}{v_T}-<F_{p_x}>_R))] \\ -\exp[\frac{-1}{k_BT}(U-\overline{V}R(JB\frac{v_{by}}{v_T}-<F_{p_x}>_R))]\} \tag{23}$$

with $\nu_c$ is the coherent oscillation frequency of the vortex bundle. Taking into account the fact that $(v_{bx}/v_T)<<1$, from the above equations, the vortex bundle flow velocity can be approximately obtained as

$$v_{by} = (\frac{\bar{V}}{R}\sqrt{\frac{T\,\Phi_0}{\pi\,B}})\,R\,\exp(\frac{-U}{k_B T})\{\exp[\frac{-\bar{V}\,R}{k_B T}(J\,B - |<F_p>_R|\cos\Theta)]$$

$$-\exp[\frac{+\bar{V}\,R}{k_B T}(J\,B - |<F_p>_R|\cos\Theta)]\} \tag{24}$$

$$v_{bx} = (\frac{\bar{V}}{R}\sqrt{\frac{T\,\Phi_0}{\pi\,B}})\,R\,\exp(\frac{-U}{k_B T})\,\{\exp[\frac{-\bar{V}\,R}{k_B T}(\frac{-J\,B\,|v_{by}|}{v_T} + |<F_p>_R|\sin\Theta)]$$

$$-\exp[\frac{+\bar{V}\,R}{k_B T}(\frac{-J\,B\,|v_{by}|}{v_T} + |<F_p>_R|\sin\Theta)]\} \tag{25}$$

where $\Theta$ is the mean angle between the random collective pinning force of the vortex bundle and positive $y$-direction measured in the counterclockwise sense. By considering the identities $\vec{E} = -\vec{v}_b \times \vec{B}$, $\rho_{xx} = E_x / J$, $\rho_{xy} = E_y / J$ together with Eq. (18) and bearing in mind that $\Theta$ is usually very small, the longitudinal and Hall resistivities can now be obtained, respectively, as follows:

$$\rho_{xx} = \frac{\bar{V}\sqrt{BT\,\Phi_0}}{J\sqrt{\pi}}\exp(\frac{-U}{k_B T})\{\exp[\frac{\bar{V}R}{k_B T}(JB - (\frac{\beta^C(T,B)}{\bar{V}})^{\frac{1}{2}})]$$

$$-\exp[\frac{-\bar{V}\,R}{k_B\,T}(J\,B - (\frac{\beta^C(T,B)}{\bar{V}})^{\frac{1}{2}})]\} \tag{26}$$

$$\rho_{xy} = \frac{-\bar{V}\sqrt{BT\,\Phi_0}}{J\sqrt{\pi}}\exp(\frac{-U}{k_B T})\{\exp[\frac{\bar{V}R}{k_B T}((\frac{\beta^C(T,B)}{\bar{V}})^{\frac{1}{2}}\bar{\alpha}\sqrt{\frac{T}{T_C - T}} - JB\frac{|v_{by}|}{v_T})]$$

$$-\exp[\frac{-\bar{V}R}{k_B T}((\frac{\beta^C(T,B)}{\bar{V}})^{\frac{1}{2}}\bar{\alpha}\sqrt{\frac{T}{T_C - T}} - JB\frac{|v_{by}|}{v_T})]\} \tag{27}$$

and

$$|v_{by}| = J\rho_{xx} / B \tag{28}$$

where we have identified the magnitude of the random average of the random collective pinning force per unit volume as $(\beta^C(T,B)/\bar{V})^{\frac{1}{2}}$ according to Eq. (5), $\bar{\alpha}$ is a proportional constant.

Taking into account the fact that the arguments in the exponential functions inside the curly bracket of Eqs. (26) and (27) are very small when the Lorentz force is close to the random collective pinning force, we finally obtain the temperature- and field-dependent longitudinal and Hall resistivities as

$$\rho_{xx} = \frac{\bar{\nu}\sqrt{B\,\Phi_0}}{J\sqrt{\pi T}} \exp\,(\frac{-U}{k_B T})\;(\frac{2\,\bar{V}\,R}{k_B})\,[J\,B-(\frac{\beta^C(T,B)}{\bar{V}})^{\frac{1}{2}})] \tag{29}$$

$$\rho_{xy} = \frac{-\bar{\nu}\sqrt{B\,\Phi_0}}{J\sqrt{\pi T}} \exp(\frac{-U}{k_B T})\;(\frac{2\bar{V}\,R}{k_B})\,[(\frac{\beta^C(T,B)}{\bar{V}})^{\frac{1}{2}}\bar{\alpha}\sqrt{\frac{T}{T_C-T}}-J\,B\frac{|v_{by}|}{v_T}] \tag{30}$$

respectively, with $|v_{by}|= J\rho_{xx}/B$.

In the following subsections we shall calculate the longitudinal and Hall resistivities for type-II superconducting bulk materials and thin films as functions of temperature and applied magnetic field, the results are then comparing with experiments.

### 5.1. Induced Longitudinal and Hall Resistivities for Type-II Superconducting Bulk Materials

Now let us concentrate on the case for type-II superconducting bulk materials, the volume $\bar{V}$ for the vortex bundle in Eqs. (29) and (30) is given as $\bar{V}=\pi\,R^2 L$, where $R\,(L)$ is the transverse (longitudinal) size of the vortex bundle. In this case, the longitudinal and Hall resistivities for type-II superconducting bulk materials now become

$$\rho_{xx} = \frac{\bar{\nu}\sqrt{\Phi_0}\sqrt{B}}{J\sqrt{\pi}\sqrt{T}}\exp(\frac{-U}{k_B T})\;[\frac{2\pi R^3 L}{k_B}][J\,B-(\frac{\beta^C(T,B)}{\bar{V}})^{\frac{1}{2}}] \tag{31}$$

$$\rho_{xy} = \frac{-\bar{\nu}\sqrt{\Phi_0}\sqrt{B}}{J\sqrt{\pi}\sqrt{T}}\exp(\frac{-U}{k_B T})\;[\frac{2\pi R^3 L}{k_B}]\,[(\frac{\beta^C(T,B)}{\bar{V}})^{\frac{1}{2}}\bar{\alpha}\sqrt{\frac{T}{T_c-T}}-J\,B\frac{|v_{by}|}{v_T}] \tag{32}$$

respectively, with $|v_{by}|= J\rho_{xx}/B$.

#### 5.1.a. Longitudinal and Hall Resistivities for Constant Temperature

Under the framework of present theory, the numerical calculations of $\rho_{xx}$ and $\rho_{xy}$ of Eqs. (31) and (32) when the temperature is kept at $T=91K$ are given in table 1. These results are in agreement with the experimental data on $YBa_2Cu_3O_{7-\delta}$ high-$T_c$ bulk materials [17]. In obtaining the above results, the following approximate data have been employed: $R=2\times10^{-8}m$, $L=10^{-6}m$, $J=10^6 A/m^2$, $T_c=92K$, $v_T=10^3 m/\sec$, $\bar{\nu}=10^{11}\sec^{-1}$, $\bar{\alpha}=5.59\times10^{-5}T^{-1/2}$, $\exp(-U/k_B\,T)=2.07\times10^{-2}m$, $(\beta^C(B=3.5)/\bar{V})^{1/2}=3.4506\times10^6 N/m^3$, $(\beta^C(3.03)/\bar{V})^{1/2}=2.9849\times10^6 N/m^3$, $(\beta^C(2.5)/\bar{V})^{1/2}=2.4605\times10^6 N/m^3$, $(\beta^C(2)/\bar{V})^{1/2}=1.9668\times10^6 N/m^3$, $(\beta^C(1.5)/\bar{V})^{1/2}=1.4748\times10^6 N/m^3$, $(\beta^C(1)/\bar{V})^{1/2}=9.854\times10^5 N/m^3$, $(\beta^C(0.75)/\bar{V})^{1/2}=7.4042\times10^5 N/m^3$, and $(\beta^C(0.5)/\bar{V})^{1/2}=4.915\times10^5 N/m^3$.

### 5.1.b. Longitudinal and Hall Resistivities for Constant Magnetic Field

Within the framework of present theory, the numerical estimations of the Hall and longitudinal resistivities when the applied magnetic field is kept at a constant value $B = 2.24$ Tesla are given in table 2. These results are in agreement with the experimental data on $YBa_2Cu_3O_{7-\delta}$ high-$T_c$ bulk materials [17]. In arriving at the above results, the following approximate data have been used: $R = 2\times10^{-8} m$ , $L = 10^{-6} m$ , $J = 10^6 A/m^2$ , $T_c = 92K$ , $v_T = 10^3 m/\sec$ , $\bar{\nu} = 10^{11}\sec^{-1}$ , $\bar{\alpha} = 5.59\times10^{-5} T^{-1/2}$ , $\exp(-U/k_B T) = 2.07\times10^{-2}$ , $(\beta^C(T = 91.6)/\bar{V})^{1/2} = 2.178\times10^6 N/m^3$ , $(\beta^C(91.3)/\bar{V})^{1/2} = 2.194\times10^6 N/m^3$, $(\beta^C(91)/\bar{V})^{1/2} = 2.204\times10^6 N/m^3$, $(\beta^C(90)/\bar{V})^{1/2} = 2.217\times10^6 N/m^3$, $(\beta^C(89)/\bar{V})^{1/2} = 2.223\times10^6 N/m^3$, and $(\beta^C(88)/\bar{V})^{1/2} = 2.224\times10^6 N/m^3$.

### 5.2. Induced Longitudinal and Hall Resistivities for Type-II Superconducting Films

Now let us turn the attention to type-II superconducting films, the volume $\bar{V}$ of the vortex bundle is therefore given by $\bar{V} = \pi R^2 d$ , where $R$ is the transverse size of the vortex bundle and $d$ is the thickness of the film. The longitudinal and Hall resistivities of Eqs. (29) and (30) can now be described, respectively, as

$$\rho_{xx} = \frac{\bar{\nu}\sqrt{\Phi_0}\sqrt{B}}{J\sqrt{\pi}\sqrt{T}}\exp(\frac{-U}{k_B T})[\frac{2\pi R^3 d}{k_B}][JB - (\frac{\beta^C(T,B)}{\bar{V}})^{\frac{1}{2}}] \quad (33)$$

$$\rho_{xy} = \frac{-\bar{\nu}\sqrt{\Phi_0}\sqrt{B}}{J\sqrt{\pi}\sqrt{T}}\exp(\frac{-U}{k_B T})\,[\frac{2\pi R^3 d}{k_B}]\,[(\frac{\beta^C(T,B)}{\bar{V}})^{\frac{1}{2}}\,\bar{\alpha}\sqrt{\frac{T}{T_c - T}} - J\,B\frac{|v_{by}|}{v_T}] \quad (34)$$

with $|v_{by}| = J\rho_{xx}/B$ .

### 5.2.a. Longitudinal and Hall Resistivities for Constant Temperature

Under the framework of the present theory, the numerical calculations of $\rho_{xy}$ and $\rho_{xx}$ as functions of applied magnetic field in Tesla when temperature is kept at a constant value $T = 4.5K$ are given in table 3. These results are in agreement with experimental data on $Mo_3Si$ conventional low-$T_c$ superconducting films [18]. In obtaining the above results, the following approximate data have been employed: $R = 2\times10^{-8} m$ , $d = 5\times10^{-8} m$ , $J = 1.5\times10^5 A/m^2$ , $T_c = 7.5K$ , $v_T = 30\, m/\sec$ , $\bar{\alpha} = 1.0449\times10^{-3} T^{-1/2}$ , $\bar{\nu} = 10^{11}\sec^{-1}$ , $\exp(-U/k_B T) = 3.0899\times10^{-4}$ , $(\beta^C(B = 7.5)/\bar{V})^{1/2} = 4.5772\times10^5 N/m^3$, $(\beta^C(7.25)/\bar{V})^{1/2} = 4.4737\times10^5 N/m^3$, $(\beta^C(7)/\bar{V})^{1/2} = 4.4324\times10^5 N/m^3$, $(\beta^C(6.75)/\bar{V})^{1/2} = 4.3964\times10^5 N/m^3$, $(\beta^C(6.5)/\bar{V})^{1/2} = 4.3483\times10^5 N/m^3$, $(\beta^C(6.25)/\bar{V})^{1/2} = 4.272\times10^5 N/m^3$, $(\beta^C(6)/\bar{V})^{1/2} = 4.0516\times10^5 N/m^3$, $(\beta^C(5.75)/\bar{V})^{1/2} = 3.7418\times10^5 N/m^3$, and $(\beta^C(5.5)/\bar{V})^{1/2} = 3.335\times10^5 N/m^3$.

### 5.2.b. Longitudinal and Hall Resistivities for Constant Magnetic Field

The numerical estimations of the $\rho_{xy}$ and $\rho_{xx}$ as functions of temperature when the applied magnetic field is kept at a constant value $B = 2$ Tesla are given in table 4. These results are in agreement with the experimental data on $Tl_2Ba_2Cu_2O_8$ high-$T_c$ superconducting films [19]. In obtaining the above results, the following approximate data have been used:

$R = 2\times10^{-8} m$, $d = 10^{-6} m$, $J = 10^7 A/m^2$, $T_c = 104K$, $v_T = 10^2 m/\sec$, $\bar{\nu} = 10^{11} \sec^{-1}$, $\bar{\alpha} = 1.12\times10^{-4} T^{-1/2}$, $\exp(-U/k_B T) = 8.3199\times10^{-5}$, $(\beta^C(T = 102)/\bar{V})^{1/2} = 1.8464\times10^7 N/m^3$, $(\beta^C(100)/\bar{V})^{1/2} = 1.9031\times10^7 N/m^3$, $(\beta^C(98)/\bar{V})^{1/2} = 1.93267\times10^7 N/m^3$, $(\beta^C(96)/\bar{V})^{1/2} = 1.9512\times10^7 N/m^3$, $(\beta^C(92)/\bar{V})^{1/2} = 1.955\times10^7 N/m^3$, $(\beta^C(88)/\bar{V})^{1/2} = 1.95618\times10^7 N/m^3$, $(\beta^C(84)/\bar{V})^{1/2} = 1.9588\times10^7 N/m^3$, $(\beta^C(78)/\bar{V})^{1/2} = 1.961069\times10^7 N/m^3$, and $(\beta^C(76)/\bar{V})^{1/2} = 1.9631\times10^7 N/m^3$.

## 6. Conclusion

The theory of the thermally activated vortex bundles flow over the directional-dependent potential barrier induced by the Magnus force, random collective pinning random force, and strong pinning force inside the vortex bundles for type-II superconductors is developed. The coherent oscillation frequency and the mean direction of the random collective pinning force of the vortex bundle are evaluated. The bundle flow velocity is obtained. Finally, the longitudinal and Hall resistivities induced by the bundle flow are calculated for type-II superconducting bulk materials as well as thin films. The results are in agreement with the experiments.

## Acknowledgement

The author would like to thank Professors E. H. Brandt and H. E. Horng for helpful discussions and suggestions.

## Tables

**Table 1.** $\rho_{xy}$ and $\rho_{xx}$ versus applied magnetic field in Tesla for $YBa_2Cu_3O_{7-\delta}$ high-$T_c$ superconducting bulk materials at $T = 91\ K$.

| $B\,(T)$ | $\rho_{xy}\ (\Omega\,m)$ | $\rho_{xx}\ (\Omega\,m)$ |
|---|---|---|
| 3.5 | $1.2914\times10^{-9}$ | $1.8739\times10^{-6}$ |
| 3.03 | $1.7658\times10^{-15}$ | $1.5916\times10^{-6}$ |
| 2.5 | $-1.4627\times10^{-9}$ | $1.2663\times10^{-6}$ |
| 2.0 | $-2.7907\times10^{-9}$ | $9.5142\times10^{-7}$ |
| 1.5 | $-3.9996\times10^{-9}$ | $6.2539\times10^{-7}$ |
| 1.0 | $-4.6405\times10^{-9}$ | $2.9669\times10^{-7}$ |
| 0.75 | $-3.9801\times10^{-9}$ | $1.6825\times10^{-7}$ |
| 0.5 | $-2.0098\times10^{-9}$ | $1.226\times10^{-7}$ |

**Table 2.** $\rho_{xy}$ and $\rho_{xx}$ versus temperature for $YBa_2Cu_3O_{7-\delta}$ high-$T_c$ superconducting bulk materials at $B = 2.24$ Tesla.

| $T\,(K)$ | $\rho_{xy}\ (\Omega\,m)$ | $\rho_{xx}\ (\Omega\,m)$ |
|---|---|---|
| 91.6 | $6.96\times10^{-10}$ | $1.878\times10^{-6}$ |
| 91.3 | $-1.82\times10^{-11}$ | $1.396\times10^{-6}$ |
| 91 | $-2.491\times10^{-9}$ | $1.094\times10^{-6}$ |
| 90 | $-4.199\times10^{-9}$ | $7.028\times10^{-7}$ |
| 89 | $-4.722\times10^{-9}$ | $5.22\times10^{-7}$ |
| 88 | $-3.081\times10^{-9}$ | $4.91\times10^{-7}$ |

**Table 3.** $\rho_{xy}$ and $\rho_{xx}$ versus applied magnetic field in Tesla for $Mo_3Si$ conventional low-$T_c$ superconducting films at $T = 4.5\ K$.

| $B\,(T)$ | $\rho_{xy}\ (\Omega\,m)$ | $\rho_{xx}\ (\Omega\,m)$ |
|---|---|---|
| 7.5 | $4.3399\times10^{-11}$ | $8.2782\times10^{-7}$ |
| 7.25 | $1.5804\times10^{-11}$ | $7.8079\times10^{-7}$ |
| 7.0 | $-2.6037\times10^{-11}$ | $7.2721\times10^{-7}$ |
| 6.75 | $-6.7189\times10^{-11}$ | $6.742\times10^{-7}$ |
| 6.5 | $-1.023\times10^{-10}$ | $6.24\times10^{-7}$ |
| 6.25 | $-1.283\times10^{-10}$ | $5.78\times10^{-7}$ |
| 6.0 | $-1.1842\times10^{-10}$ | $5.492\times10^{-7}$ |
| 5.75 | $-8.8118\times10^{-11}$ | $5.3044\times10^{-7}$ |
| 5.5 | $-3.752\times10^{-11}$ | $5.2209\times10^{-7}$ |

**Table 4.** $\rho_{xy}$ and $\rho_{xx}$ as functions of temperature for $Tl_2Ba_2Cu_2O_8$ high-$T_c$ superconducting thin films at $B = 2$ Tesla.

| $T\ (K)$ | $\rho_{xy}\ (\Omega\, m)$ | $\rho_{xx}\ (\Omega\, m)$ |
|---|---|---|
| 102 | $2.1346\times10^{-11}$ | $1.6728\times10^{-8}$ |
| 100 | $-3.69\times10^{-19}$ | $1.0657\times10^{-8}$ |
| 98 | $-1.4607\times10^{-11}$ | $7.4819\times10^{-9}$ |
| 96 | $-2.3518\times10^{-11}$ | $5.4754\times10^{-9}$ |
| 92 | $-1.0344\times10^{-11}$ | $5.1609\times10^{-9}$ |
| 88 | $5.9409\times10^{-19}$ | $5.1382\times10^{-9}$ |
| 84 | $5.46\times10^{-12}$ | $4.951\times10^{-9}$ |
| 78 | $1.3011\times10^{-11}$ | $4.8488\times10^{-9}$ |
| 76 | $1.2967\times10^{-11}$ | $4.65\times10^{-9}$ |